\documentclass[12pt]{iopart}
\usepackage{epsf}

\begin{document}
%\maketitle
\title{Moving Beyond a Simple Model of Luminescence Rings in Quantum Well
Structures}
\author{D. Snoke$^1$, S. Denev$^1$, Y. Liu$^1$, S. Simon$^2$, R. Rapaport$^2$, G.
Chen$^2$, L. Pfeiffer$^2$, K. West$^2$}
\address{$^1$University of Pittsburgh, 3941 O'Hara St., Pittsburgh, PA 15260, USA\\
$^2$Bell Labs, Lucent Technologies, 700 Mountain Ave., Murray Hill, NJ 07974-0636, USA}

\begin{abstract}
The dramatic appearance of luminescence rings with radius of several hundred microns
in quantum well structures can be understood through a fairly simple nonlinear model
of the diffusion and recombination of electrons and holes in a driven
nonequilibrium system. The ring corresponds to the boundary between a positive hole
gas and a negative electron gas in steady state. While this basic effect is now well
understood, we discuss several other experimental results which can not be explained by
this simple model.
\end{abstract}
\pacs{71.35.Lk, 71.10.Ca,  73.63.Hs, 78.67.De}

%\newpage
\section{Introduction}
In August of 2002, two articles in Nature magazine \cite{snoke-nature,butov-nature}
presented a new effect--- sharply-defined rings of luminescence from quantum well
structures with radius of several millimeters around a much smaller,
central laser spot. Figure 1 shows the basic effect. The Berkeley group
\cite{butov-nature} also reported breakup of the rings under certain conditions into a
periodic array of bright spots. 
\begin{figure}[htpb]
\begin{center}
\hspace{2cm}
\epsfxsize=.7\hsize 
\epsfbox{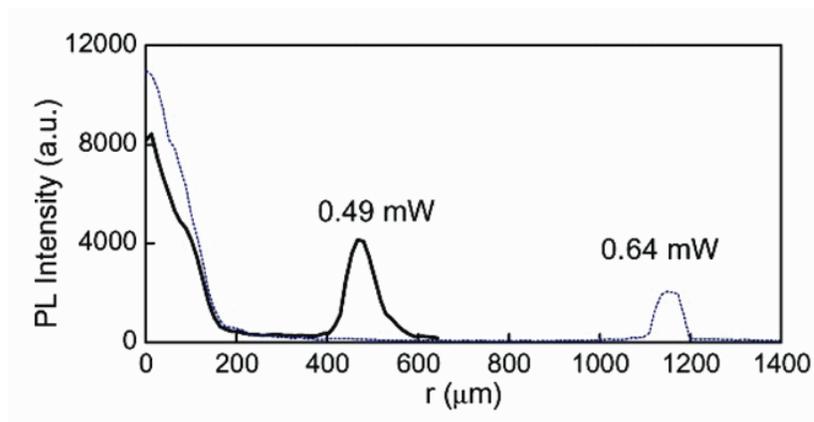} 
\end{center}
%\picture 6in by 6in (fig1 scaled 900)
\caption{Spatial profiles of the luminescence from Structure A (see \protect\ref{tab1}), for two
different laser powers, with applied voltage $3.3$ V, at $T=2$ K. The excitation laser was a
continuous HeNe laser at 632.8 nm.}
\end{figure}

The basic mechanism for the formation of the rings is now well
understood. As described in earlier papers \cite{ssc,noeks}, several experimental
facts point toward a consistent explanation. The most important is that the ring only
appears when the laser photon energy is large enough to cause hot carriers to jump
over the outer barriers of the quantum wells. Figure 2 shows a comparison of the
efficiency of ring creation for various samples. As seen by comparing the results from
structures A and C, the photon energy needed for creation of the ring follows the
height of the outer barriers. Also, the ring effect only occurs in samples with
$n-i-n$ doping, in which an $n$-type conducting layer is on the other side of the
barriers. Another key observation is that the ring can be created in single quantum
wells as well as in double quantum wells with zero applied voltage, which means that
the lifetime of the excitons is not a major factor.
\begin{figure}[htpb]
\begin{center}
\hspace{2cm}
\epsfxsize=.7\hsize 
\epsfbox{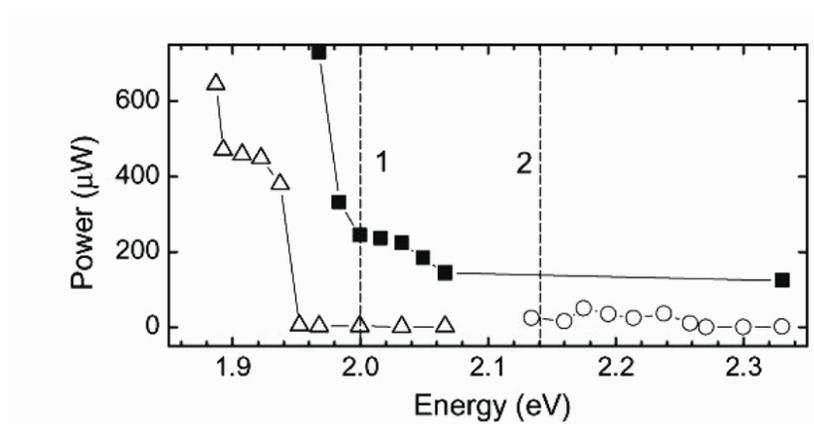} 
\end{center}
%\picture 4in by 4in (fig2 scaled 700)
\vspace{1cm}
\caption{Critical laser power for the ring effect as a function of laser photon
energy, for three different structures at $T=2 $K. Below this critical density,
the ring does not exist, because its radius decreases with power until it merges
with the central luminescence spot at the critical density.  Open
triangles: Structure A (see Table 1). Solid squares: Structure B.  This
structure reproduces the structure design reported in Ref. \protect\cite{butov-nature}.
No ring is seen for either Structure A or B for photon energy less than 1.85 eV. The
vertical dashed line at 2.0 eV gives the outer barrier height for these structures.
Open circles: Structure C. This is the same as structure A, but with 
Al$_{0.44}$Ga$_{0.56}$As outer barriers (direct gap shown by the vertical line at 2.14
eV.) No ring is seen in structure C for photon energy less than 2.1 eV. Gap energies
are taken from Ref.
\protect\cite{adachi}.}
\end{figure}

These facts plus
several other results are consistent with the following picture: when the laser
creates hot carriers in the quantum well, some of the holes and
electrons jump the outer barriers and leave the quantum well.  If this is
slightly more likely for electrons than holes, then an excess population
of holes will build up near the laser excitation spot. At
the same time, electrons from the outer $n$-doped layers tunnel into the
quantum well through the barriers at a very slow, but constant, rate. Therefore, far
from the laser spot,  the wells will have an excess of electrons. The luminescence
ring is the steady-state boundary between the hole-rich region centered on the laser
excitation spot and the electron-rich region outside.
\begin{table}
\begin{center}
\begin{tabular}{l|ll}
sample & description\\ \hline
A & two 60 \AA \ In$_{0.1}$Ga$_{0.9}$As wells separated by 40 \AA \ GaAs
barrier\\
& 300 \AA \ and 1000 \AA \ Al$_{0.32}$Ga$_{0.68}$As outer barriers
\\   \hline
B & two 80 \AA \ wells separated by 40 \AA~ GaAs barrier\\
& 2000 \AA \ Al$_{0.33}$Ga$_{0.67}$As outer barriers \\ \hline
C & two 60 \AA \ In$_{0.1}$Ga$_{0.9}$As wells separated by 40 \AA \
GaAs barrier\\ 
& 300 \AA \ and 1000 \AA \ Al$_{0.45}$Ga$_{0.55}$As outer barriers\\ \hline
D & single 60 \AA \ In$_{0.1}$Ga$_{0.9}$As well\\ 
& 300 \AA \ and 1000 \AA \ Al$_{0.32}$Ga$_{0.68}$As outer barriers
\end{tabular}
\end{center}
\caption{Samples used in this study. All are $n-i-n$ structures with heavily
$n$-doped GaAs on either side of the outer barriers.}
\label{tab1}
\end{table}

This effect has been successfully modeled numerically, in a model
presented in Ref. \cite{condmat}. A similar model has also been used by
Butov et al. \cite{lev}.  Figure 3 shows the basic result of the
model.  
\begin{figure}[htpb]
\begin{center}
\hspace{2cm}
\epsfxsize=.7\hsize 
\epsfbox{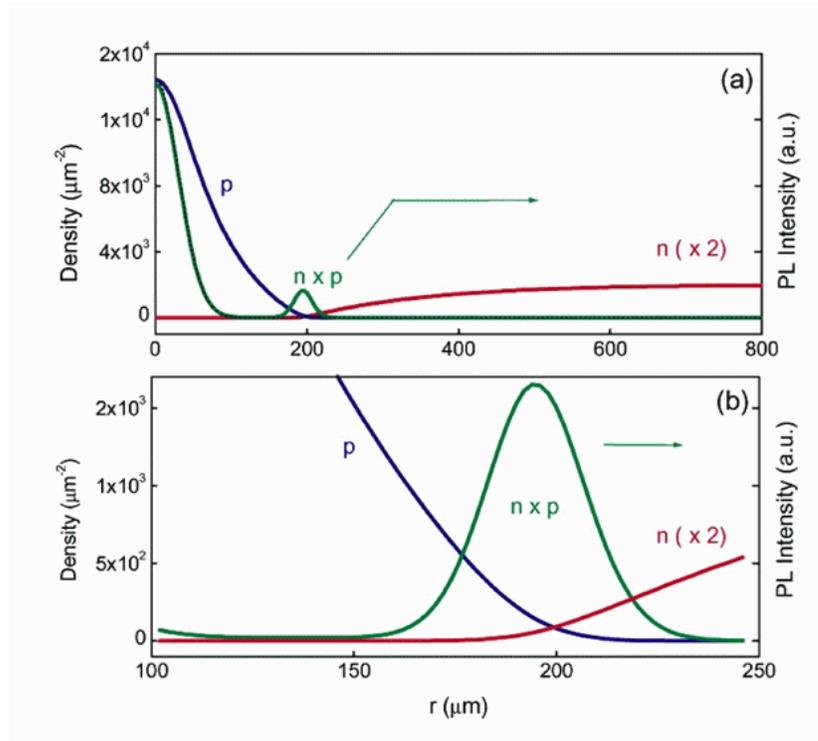} 
\end{center}
%\picture 6in by 6in (fig3 scaled 900)
\caption{a) Prediction of the
theoretical model of Ref.
\protect\cite{condmat} for the luminescence intensity profile. The curves marked {\sf p} and {\sf n}
give the distribution of holes and electrons in steady state, respectively, while the curve marked
{\sf n}$\times${\sf p} gives the relative luminescence intensity. The plot of (b) is an expanded plot
of the same theory shown in (a).}
\end{figure}

The model of Ref. \cite{condmat} uses a set of nonlinear equations which reduce
to the following:
\begin{eqnarray}
\frac{\partial n}{\partial t} &=& D_e\nabla^2 n  - Anp +
G_e(\vec{r}) - \frac{n-n_0}{\tau} 
\\
\frac{\partial p}{\partial t} &=& D_h\nabla^2 p -   Anp + G_h(\vec{r}),
\end{eqnarray}
where $n$ and $p$ are the density of electrons and holes in the well structure,
respectively, and they diffuse in the plane of the well structure with diffusion
constants $D_e$ and $D_h$. Electrons and holes are
both generated by the laser pulse near $\vec{r} = 0$, with slightly different
rates $G_e$ and $G_h$, since some hot electrons hop the outer barriers and do
not enter the wells.  In addition, electrons are either lost or
gained by tunneling through the outer barriers, depending on the density of electrons
relative to the equilibrium value
$n_0$ (in the Berkeley experiments, this tunneling only occurs at high voltage
when the quantum well levels lie below the GaAs bulk electron states).  The nonlinear
term $Anp$ represents the rate of exciton formation which leads to recombination of
the carriers with a rate proportional to the product of $n$ and $p$. As discussed in
\cite{condmat}, the sharpness of the ring, which is the most surprising effect, arises
from the nonlinearities introduced by this term.

In this paper, we concentrate on some of the experimental features not
yet explained by the model. Although the model presented above 
is in good quantitative agreement with several aspects of the
experimental observations, there are some effects that still cannot be
accounted for.

\section{Dependence of the Ring Radius on Excitation Density}
 As seen in
Figure 4, although a ring can be created in both double and single quantum well
structures, for the double quantum well structure for certain experimental parameters,
the ring gets bigger with decreasing laser spot size while laser power is held
constant. For the single quantum well sample, the ring gets smaller with decreasing
spot size for all experimental conditions.  In the numerical model, the ring gets
smaller with decreasing spot size for all values of the parameters; the model does not
take into account whether the sample is a double or single quantum well, since it
keeps track only of the total particle density in the structure.

One possible explanation for the increase of the ring radius with decreasing
spot size is that the fraction of electrons which jump the barrier becomes
greater when the laser focus is tighter. In the numerical model, the ratio of
electrons to holes generated by the laser in the quantum well structure is
assumed constant. If the laser focus is very tight, however, then the
temperature of the electrons can become elevated, leading to more of them
jumping the outer barriers. It is not clear why this should only occur in the double
quantum well sample; possibly the rate of cooling is more efficient in the
single quantum well.

Another effect which is not well understood is the non-monotonic dependence of the
ring radius on laser power when the laser power is well above the critical threshold.
At very low excitation density, there is no ring. As discussed earlier
\cite{snoke-nature,ssc}, when the density passes a critical threshold, the ring
appears and grows larger as the carrier density increases. As the laser power
continues to increase, however, the ring does not continue to grow indefinitely. As
shown in Figure 5, the ring radius can decrease and then increase again; in some
cases there is more than one maximum radius as the ring radius appears to go through
oscillations. 

This effect may also be understood in terms of competition between the excitation
which leads to accumulation of carriers in the wells and the heating processes which
affect the jumping of carriers over the barriers. As the density of
carriers in the well increases, at some point the probability of a second photon from
the laser exciting these carriers to high energy will become significant. At this
point the relative rates of holes and electrons jumping the barriers will become more
equal, since highly excited electrons and holes will both have energy greatly in
excess of the barrier height.  Since the effect of the ring depends crucially on the
difference of the rate of electrons and holes jumping over the outer barriers, even
small changes of the relative rates may lead to large changes in the steady state
population of holes, and therefore the ring radius.

\section{Other Experimental Effects}

Although we have reproduced the design of the sample used by the Berkeley group, there
are two effects reported by them which we have not reproduced. Differences in the
experiments include the possibility of different amounts of disorder (the sample used
by the Berkeley group had fixed, bright spots which could correspond to
defects) and different spatial resolution (our resolution is approximately 5 $\mu$m.)

One effect seen by the Berkeley group, which is not reproduced by the numerical model
of Ref. \cite{condmat}, is a small, ``inner ring''. Although
we do not see a clearly-defined inner ring, we do observe a slight shoulder on the
inner spot in several cases, as seen in Figure 1, which can be interpreted as a ring
near the central spot, convolved with the spatial resolution of the imaging system. 
Ivanov has argued \cite{ivanov} that this inner ring can be understood through a
kinetic effect of quasi-ballistic scattering of hot carriers near the laser generation
spot. In this picture, the excitons remain at high energy for a short period of time,
during which they are outside the region of momentum space in which they can emit
photons. Scattering tends to equilibrate the population of excitons into the low energy
states which couple to photons, so that after they have traveled a short distance,
they cool down into these states and emit light, giving rise to the extra ring.

Another, dramatic effect is the breakup of the ring at low temperature, which also is
not explained by the numerical model discussed above.  We have
also not yet been able to reproduce this result experimentally, although as discussed
above, we have copied the published design of the AlGaAs/GaAs structure used in those
experiments
\cite{butov-nature}, and we have used similar experimental conditions, i.e. $T=1.8$ K,
to those reported by the Berkeley group. 

Levitov and coworkers \cite{lev} have given one possible explanation for this effect.
Another possibility for the breakup of the ring into a periodic pattern relates to the
intrinsic instability of the boundary between the electron-rich and hole-rich regions. 
The Coulomb repulsion of carriers with like charge and the Coulomb attraction of
carriers with opposite charge makes the holes want to move outward and the
electrons want to move inward. In the normal case, carriers do not penetrate far
into the region of opposite charge because they annihilate by photon emission as soon
as they do. This is why the luminescence ring is so sharply defined. At low
temperature, however, the boundary may become unstable as the mobility of the carriers
increases.
\begin{figure}[htpb]
\begin{center}
\vspace{1cm}
\hspace{2cm}
\epsfxsize=.6\hsize 
\epsfbox{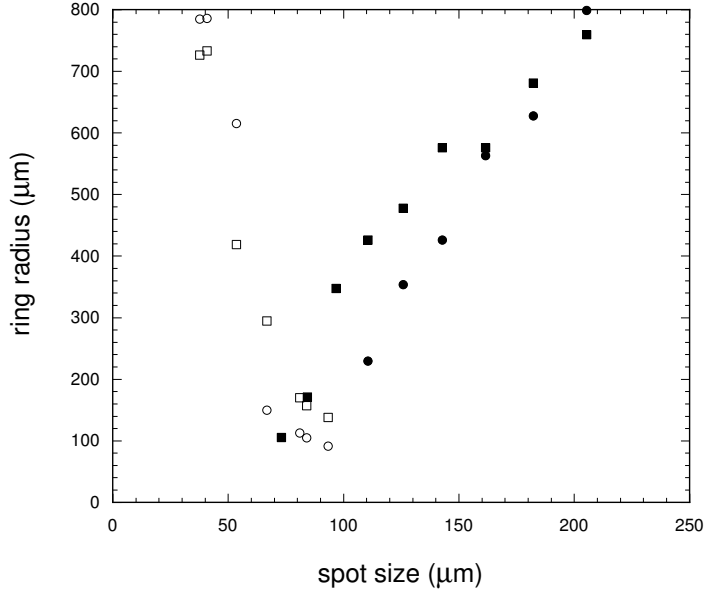} 
\end{center}
%\picture 6in by 6in (fig4 scaled 900)
\vspace{1cm}
\caption{Laser focus dependence of the ring radius, for continuous laser
excitation at 632.8 nm with average power 4 mW. Open squares: coupled quantum well,
$T=$ 15 K. Open circles: coupled quantum well (sample A), $T=$ 2 K.  Filled circles:
single quantum well (sample D), $T=2$ K. Filled squares: single quantum well, $T=15 $K.}
\end{figure}

\begin{figure}[htpb]
\begin{center}
\vspace{1cm}
\hspace{2cm}
\epsfxsize=.6\hsize 
\epsfbox{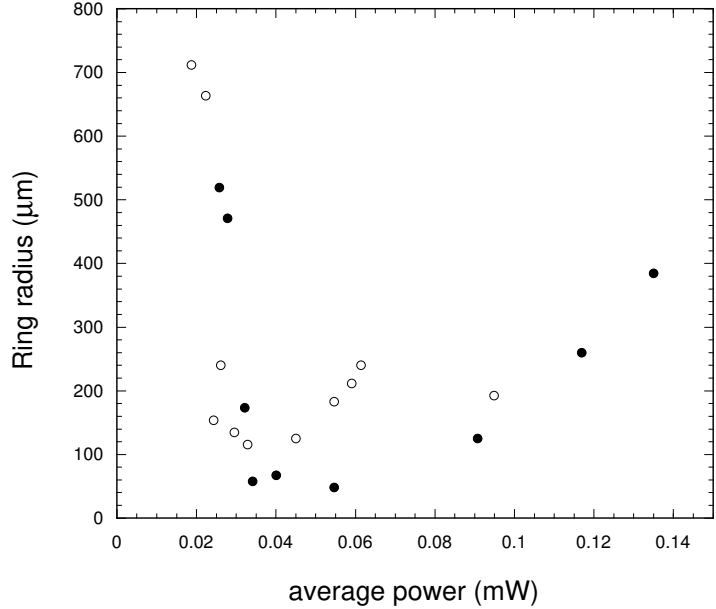} 
\end{center}
%\picture 6in by 6in (fig5 scaled 900)
\vspace{1cm}
\caption{Power dependence of the ring radius, for Sample A. Open circles: $T=$ 4K,
applied voltage $3.43 $V.
 Filled circles: $T=$ 30 K, applied voltage $3.79$ V. }
\end{figure}

\section{A Model of Screened Coulomb Pressure}

Although the hole-rich and electron-rich regions consist of just one type of charge,
there is not a long-range Coulomb force in this system. On the outside of the barriers
the heavily-doped $n$-type GaAs acts as a conductor which will screen out electric
field on length scales long compared to the outer barrier thickness, which is around
1000 \AA, or 0.1 $\mu$m. The proper model for this system is therefore a gas with
short range Coulomb drift force. 

We can write down an altered transport equation
taking into account this force, using the standard theory. 
In general, the average velocity of electrons in a system with
drift and diffusion is
\begin{equation}
\vec{v} = -D\frac{\nabla n}{n} - {\mu}\vec{E},
\end{equation}
where $\mu$ is the electron mobility. Multiplying by $n$ and taking the divergence of
both sides, we have
\begin{equation}
\nabla\cdot (n\vec{v}) = -D{\nabla^2 n} - {\mu}\nabla\cdot(\vec{E}n)
\end{equation}
By the continuity equation, $\partial n/\partial t  + \nabla\cdot(n\vec{v}) = 0$, we
therefore have
\begin{eqnarray}
\frac{\partial n}{\partial t} &=& D\nabla^2 n +\mu \nabla\cdot(\vec{E}n) .
%\\ &=& D\nabla^2 n +\mu (n\nabla\cdot\vec{E}+\vec{E}\cdot\nabla n). \nonumber
\end{eqnarray}
For a charge screened by a conducting plane at a distance $d$, we have everywhere locally,
\begin{equation}
\vec{E} = \frac{4\pi e d}{\epsilon}\nabla(n-p),
\end{equation}
where $e$ is the fundamental charge.
If we assume that the gradient of $n$ is small, i.e. features are large compared to
the screening length, Equations (1) and (2) become
\begin{eqnarray}
\frac{\partial n}{\partial t} &=& D_e\nabla^2 n +\frac{4\pi \mu_e e d}{\epsilon}
{\nabla\cdot n\nabla(n-p)}  - Anp + G_e(\vec{r}) - \frac{n-n_0}{\tau} \\
\frac{\partial p}{\partial t} &=& D_h\nabla^2 p +\frac{4\pi \mu_h e d}{\epsilon}
{\nabla\cdot p\nabla(p-n)}  -   Anp + G_h(\vec{r}) .
\end{eqnarray}
These nonlinear equations may lead to all manner of
pattern forming behavior. Since for a degenerate gas, $D = v_F^2\tau$ is proportional to the carrier
density, an additional alteration to the above equations is to include the density dependence of $D$,
in which case the term $\nabla^2 n$ should be replaced by $\nabla\cdot(D\nabla n) =
C \nabla\cdot (n\nabla n)$, where $C=2\pi\hbar^2\tau/m^2$. In this case,  both the diffusion term and
the Coulomb pressure term have the same form. A simple estimate indicates that the Coulomb pressure
term dominates over the standard diffusion term at all densities. This indicates that the simple model
of Equations (1) and (2) should be adjusted to give more realistic account of the transport.

Unfortunately, the revised equations (7) and (8) are much less tractable to numerical solution than
Equations (1) and (2). Terms of the form ${\nabla\cdot n\nabla(p-n)}$ which involve a difference
between the two densities $p$ and $n$ are very sensitive to small changes, since it is well known in
numerical mathematics that subtraction of two comparable numbers leads to an increase of relative
error. Work to solve these equations for realistic parameters is still in progress.

\section{Conclusions}
As we have seen, the model of Ref. \cite{condmat} adequately reproduces the basic
effect of the ring, but it leaves out two effects, namely, the dependence of the
generation rates $G_e$ and $G_h$ on the excitation density, and the short-range Coulomb
interaction of the carriers. At present, there are several experimental results which
cannot be accounted for by this simple model, and  these may be
related to the extra complexities which these additional effects introduce.

{\bf Acknowledgements}. This work has been supported by the National Science Foundation under
Grant No. DMR-0102457 and by the Department of Energy under Grant No. DE-FG02-99ER45780.

\end{document}